\documentclass{article}
\usepackage{amsmath,amssymb,epsfig}

\newcommand{\beq}{\begin{equation}}
\newcommand{\eeq}{\end{equation}}

\begin{document}

\thispagestyle{empty}
\begin{flushright} HU-EP-19/22
\end{flushright}

\vskip 2 cm

\begin{center}
{\huge \textbf{Polylogarithms from the bound state $S$-matrix}}

\vskip 1 cm

M.~de Leeuw, Trinity College Dublin, \\
B.~Eden, D.~le Plat and T.~Meier, Humboldt-University Berlin, 
\vskip 1.5 cm

To be contributed to the proceedings of the LT13 conference at the Bulgarian Academy of Sciences, Varna, the 17th to the 23rd of June, 2019. 
\end{center}

\vskip 2 cm

Higher-point functions of gauge invariant composite operators in ${\cal N}=4$ super Yang-Mills theory can be computed via triangulation. The elementary tile in this process is the hexagon introduced for the evaluation of structure constants. A glueing procedure welding the tiles back together is needed to return to the original object. 

In this note we present work in progress on $n$-point functions of BPS operators. In this case, quantum corrections are entirely carried by the glueing procedure. The lowest non-elementary process is the glueing of three adjacent tiles by the exchange of two single magnons.

This problem has been analysed before. With a view to resolving some conceptional questions and to generalising to higher processes we are trying to develop an algorithmic approach using the representation of hypergeometric sums as integrals over Euler kernels.

\newpage

\section{Introductory remarks}

The spectrum problem of the AdS/CFT correspondence in the original form --- so connecting ${\cal N} = 4$ super Yang-Mills theory in four dimensions to IIB string theory on AdS$_5\times$S$^5$ --- has been successfully described by an integrable system \cite{beiStau}. The effects of higher-loop corrections in the field theory can be incorporated into the corresponding Bethe equations using the Zhukowski variables $x(u)$ defined by
\beq
x + \frac{g^2}{2 \, x} \, = \, u \label{defXRoot}
\eeq
where $u$ is a Bethe rapidity. In particular, since the Bethe ansatz involves the quantities $u^\pm \, = \, u \pm \frac{i}{2}$ one introduces $x^\pm(u) \, = \, x(u^\pm)$. 

A double Wick rotation from the original model to a \emph{mirror theory} enables one to use the thermodynamic Bethe ansatz (TBA) for the discussion of finite size effects in the AdS/CFT integrable model \cite{zamEtc}. W.r.t. to the Bethe rapidities, the mirror transformation is (here the scaling is adapted to the string side)
\beq
\gamma \quad : \quad x^+ \, \rightarrow \, \frac{1}{x^+}
\eeq
while $x^-$ stays inert. More generally we may define \cite{BKV}
\begin{eqnarray}
2 \gamma & : & x^\pm \, \rightarrow \, \frac{1}{x^\pm} \, , \nonumber \\
3 \gamma & : & x^+ \, \rightarrow \, x^+, \ x^- \, \rightarrow \, \frac{1}{x^-} \, ,  \\
4 \gamma & : & x^\pm \, \rightarrow \, x^\pm \, . \nonumber
\end{eqnarray}
The $2 \gamma$ transformation has the interpretation of \emph{crossing} from particles to antiparticles. Other $n \gamma$ transformations are the same modulo 4 on expressions only depending on the square root functions $x^\pm$. However, another element of the integrable system is the \emph{dressing phase} \cite{BES} obeying a crossing equation \cite{janik} implying that it does not return to itself at $4 \gamma$. Developing the TBA has necessitated understanding the scattering of bound states of the model \cite{dressCross,glebBound}. 

For a long time, higher-point functions remained hard to address using these methods. The introduction of the \emph{hexagon operator} meant a break-through w.r.t. the three-point problem \cite{BKV}. In a nutshell, to evaluate the hexagon one can move all excitations to one spin chain by approriate $n \gamma$ transformations and then scatter by the $psu(2|2)$ invariant $S$-matrix \cite{psu22}, or its bound state variant \cite{glebBound}. Every scattering is accompanied by a certain scalar factor $h$ containing the (inverse of the) dressing phase and some factor of $x^\pm$ type.

Finally, it was noticed in \cite{cushions,shotaThiago1} that higher-point functions can likely be evaluated by tilings with hexagon patches, for a four-point function see Figure 1. The circular openings in the figure are spin chains equivalent to the gauge theory operator. The faces of the figure yield four hexagons. The cut in the left panel is not promising because it introduces a sum over a complete set of physical states (the OPE), already at tree level. The second cut is suggested by tree-level Feynman diagrams and is much more useful because there is no sum over intermediate states.
\begin{figure}[h]
\hskip 2.1 cm \includegraphics[width = 8 cm]{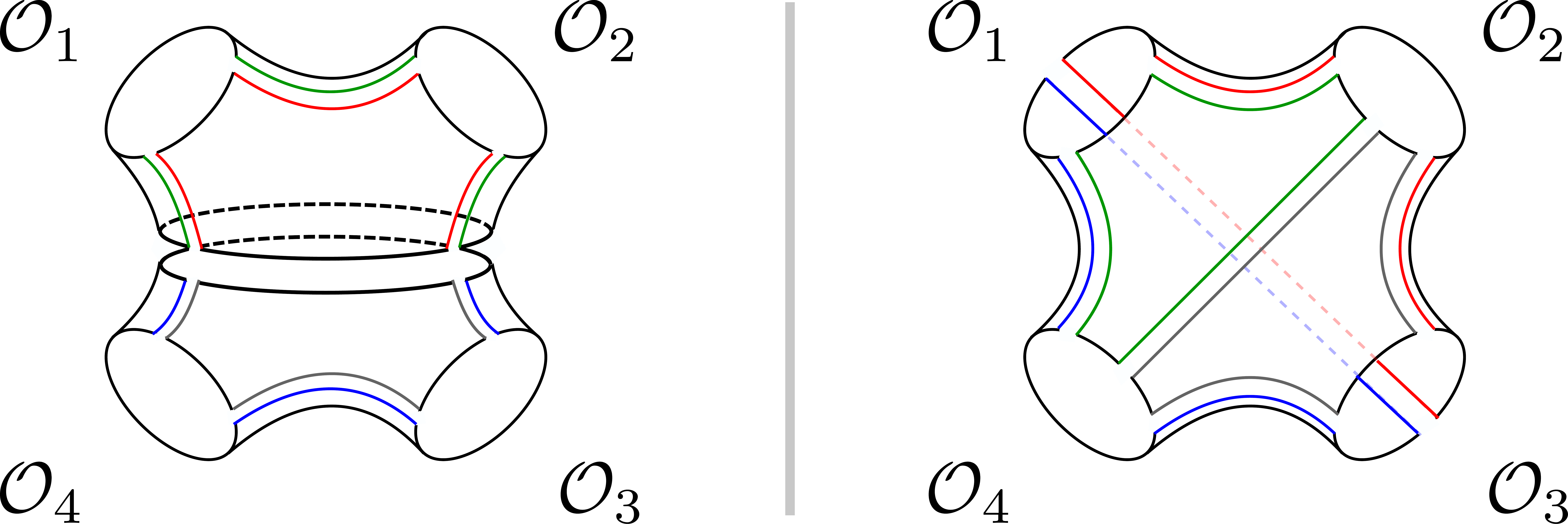}
\caption{OPE and non-OPE cuts of a four-point function}
\end{figure}
To get back to the original uncut figure the tiles are \emph{glued} by the procedure defined in \cite{BKV} for the three-point case. This can be thought of as the insertion of a complete set of states. In fact, relevant are the bound states of the TBA analysis. This is a complicated but --- as we shall see --- hopefully manageable sum.

The glueing of three adjacent tiles
\begin{figure}[h]
\hskip 4 cm \includegraphics[width=0.35\linewidth]{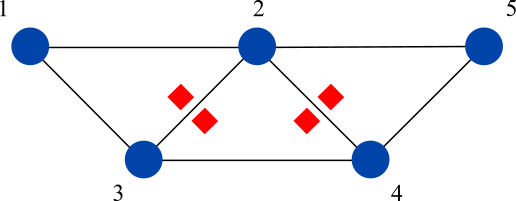}
\caption{Glueing three adjacent tiles by two virtual exchanges; the conjugate pairs of bound states are marked as red squares on the two sides of the glued edges.}
\end{figure}
has been evaluated in \cite{shotaThiago2} by matching a truncated residue calculation on an ansatz. The motivation for our study is to expand on this work: how can we integrate/sum in closed form? Second, in \cite{shotaThiago2} it became apparent that extra braiding factors $e^{i p/2}$ (here $p$ denotes the momentum of the bound state particles) have to be introduced. We would eventually like to answer whether the choice of braiding adopted there is the only possible one.

Now, according to \cite{BKV} we have to choose $sl(2)$ sector bound states, c.f. equation (\ref{sl2BoundBits}). The scattering matrix available in the literature \cite{glebBound} is originally written for the opposite case: $su(2)$ bound states obtained from (\ref{sl2BoundBits}) by exchanging the r\^ole of bosonic and fermionic constituents. We argue below that --- at least in the situation at hand --- the formulae of \cite{glebBound} apply directly.
We find a very clean integration scheme, although we cannot yet answer whether the result matches complete correlation functions.

\section{Elements of the calculation}

Let us first consider glueing two neighbouring hexagons by a single mirror magnon. Let the first hexagon, as in the three-point problem in \cite{BKV}, connect operators at the positions
\beq
x_1 \, = \, \{ 0, 1, 0, 0 \} \, , \quad x_2 \, = \, \{ 0,0,0,0 \} \, , \quad x_3 \, = \, \{0,\infty,0,0 \} \, .
\eeq
The second hexagon shares the edge between $0,\infty$ but its third point is parametrised by the variables $z,\bar z$, so it depends on
\beq
\quad x_2 \, = \, \{ 0,0,0,0 \} \, , \quad x_3 \, = \, \{0,\infty,0,0 \} \, , \quad x_4 \, = \, \{ 0,-\Im(z), \Re(z), 0 \} 
\eeq
and the glueing is over the common edge 23. In a four-point problem this is not a restriction on the kinematics. One finds the frequently used parametrisation
\beq
\frac{x_{13}^2 x_{24}^2} {x_{12}^2 x_{34}^2} \, = \, z \bar z \, , \qquad \frac{x_{14}^2 x_{23}^2}{x_{12}^2 x_{34}^2} \, = \, (1-z)(1-\bar z)
\eeq
for the two independent cross-ratios. In \cite{shotaThiago1} it is suggested to obtain the non-standard coordinates $z,\bar z$ from the usual situation $0,1,\infty$ by the operator
\beq
W(z,\bar z) \, = \, e^{-D \, \log|z|} \sqrt{\frac{z}{\bar z}}^{L} \, , \qquad L \, = L^1_1 - L^2_2 \label{rotateHex}
\eeq
where $D$ is the dilatation generator. This leaves $0,\infty$ invariant but maps $1 \, \mapsto \, (z,\bar z)$. In calculations one will not transform coordinates but rather act on the states scattering over the second hexagon. Since the operator above is diagonal on these one can finally evaluate both hexagons as in the three-point problem.

Glueing means inserting $sl(2)$ bound states \cite{BKV}. This \emph{antisymmetric representation} at level (or length) $a$ has the parts
\beq
(\psi^1)^{a-k-1} (\psi^2)^k \phi^i \, , \qquad (\psi^1)^{a-k} (\psi^2)^k \, , \qquad (\psi^1)^{a-k-1} (\psi^2)^{k-1} \phi^1 \phi^2 \, . \label{sl2BoundBits}
\eeq
Customarily, in the first case one separately considers $i \, = \, 1,2$. In a four-point calculation one can \cite{shotaThiago1} act on the \emph{states} on the second hexagon by the tilting transformation (\ref{rotateHex}). To this end we rewrite
\beq
\frac{1}{2} (D-J) \, = \, E \, = i \tilde p \, = \, i \, u + \ldots
\eeq
where the dots indicate $O(g^2)$ corrections in a weak coupling expansion. The generator $J$ acts on the scalars in the $sl(2)$ bound states, of which there are only one or two. Our purpose in this note is to re-sum the infinite series in $z,\bar z$ that the weight factor creates; for now we turn a blind eye on all transformations required to rotate the second hexagon in the internal space. We can also send $\psi^{a-k-1} \, \rightarrow \, \psi^{a-k}$ etc. since these are constant shifts, while the summation ranges must, of course, be respected to obtain sensible results.

Then,
\beq
W(z,\bar z) \, (\psi^1)^{a-k} (\psi^2)^k \, = \, (z \bar z)^{-i \, u} \left(\frac{z}{\bar z}\right)^{\frac{a}{2} - k} \, (\psi^1)^{a-k} (\psi^2)^k
\eeq
because $L$ (the Cartan generator of the Lorentz transformation) attributes weight $1,-1$ to $\psi^1, \psi^2$, respectively.

In the five-point process in Figure 2, the central tile is glued to two neighbouring hexagons. Full fledged five-point kinematics cannot be parametrised using only the coordinates of the $1,2$ plane, so that the Cartan generators used above are not enough to recover it. Nonetheless, to get started we follow \cite{shotaThiago2} and use restricted kinematics. We then obtain a weight factor for either glueing, so $W(z_1,\bar z_1) \, W(z_2,\bar z_2)$. Clearly, the fifth cross ratio is lost.

On the left and the right hexagon there is only one bound state and thus no scattering. Yet, the contraction rule for the outer hexagons enforces the scattering on the middle tile to be diagonal. Further, let us choose $3 \gamma, \, 1 \gamma$ kinematics on the middle hexagon in which case the scalar factor becomes
\beq
h(u^{3 \gamma}, v^\gamma) \, = \, \Sigma(u^\gamma, v^\gamma)
\eeq
with the \emph{improved BES dressing phase} \cite{BES,dressCross} in mirror/mirror kinematics
\beq
\Sigma^{ab} \, = \,  \frac{\Gamma[1 + \frac{a}{2} + i \, u]}{\Gamma[1 + \frac{a}{2} - i \, u]} \, \frac{\Gamma[1 + \frac{b}{2} - i \, v]}{\Gamma[1 + \frac{b}{2} + i \, v]} \, \frac{\Gamma[1 + \frac{a+b}{2} - i (u-v)]}{\Gamma[1 + \frac{a+b}{2} + i (u-v)]} \, + O(g) \, . \label{sig11}
\eeq

A comprehensive discussion of the bound state $S$-matrix is given in \cite{glebBound}. By way of example, we consider the scattering of two states of the first type in (\ref{sl2BoundBits}), both with $i=1$ or both with $i=2$. The relevant scattering matrix is called $X^{kl}_n(a,u,b,v)$ in \cite{glebBound}, where we associate the bound state counter $a$ and the rapidity $u$ as well as $x^\pm(u)$ with the first particle and $b,v,y^\pm(v)$ with the second. 

In the symmetric representation, the r\^ole of bosons and fermions is exchanged, in particular in (\ref{sl2BoundBits}). As a consequence, at bound state length 1 (so for fundamental particles) the $X$ element describes the scattering of two equal fermions. Hence in \cite{glebBound} it is called $D$ in agreement with the nomenclature of \cite{psu22}. The entire $S$-matrix can be changed by an overall factor, and indeed this $D$ is equal to the $A$-element in \cite{psu22}\footnote{It follows from here that the bound state length 1 part of the $S$-matrix of \cite{glebBound} is in fact the inverse of that derived in \cite{psu22}.}.

We repeated the steps of \cite{glebBound} to re-derive the $S$-matrix in the antisymmetric representation. Flipping the statistics means exchanging Poincar\'e and conformal supersymmetry, and also Lorentz and internal symmetry generators. Sticking to the same algebra conventions one obtains a sign flip on the rapidity parameters, so in particular $x^\pm \, \leftrightarrow \, x^\mp$. The $X$ element at bound state length 1 now describes the scattering of two equal bosons. We observe that what was called $D$ before now becomes $A^{-1}$. Hence for the antisymmetric representation the construction yields $S^{-1}$ without any rescaling.

Next, by observation --- at least in $3\gamma, \, 1\gamma$ kinematics and at leading order in $g$ --- the diagonal elements of our $S^{-1}$ in the antisymmetric representation are related to those of \cite{glebBound} by flipping the sign of the rapidities, which has the interpretation of a complex conjugation or of taking a second inverse. It follows that we can use the $S$-matrix of \cite{glebBound} for our purposes, without any changes! 

Verbatim,
\begin{eqnarray}
&& X_n^{k,l} \, = \, D \ \frac{\prod_{j=1}^n (a-j) \, \prod_{j=1}^{k+l-n} (b-j)}{\prod_{j=1}^k (a-j) \, \prod_{j=1}^l (b-j) \, \prod_{j=1}^{k+l} (-i \, \delta u + \frac{a+b}{2} - j)} \, * \nonumber \\
&& \sum_{m=0}^k \left(\begin{array}{c} k \\ k-m \end{array} \right) \left(\begin{array}{c} l \\ n-m \end{array} \right) \, \prod_{j=1}^m c^+_j \, \prod_{j=1-m}^{l-n} c^-_j \, \prod_{j=1}^{k-m} d_{k-j+2} \, \prod_{j=1}^{n-m} \tilde d_{k+l-m-j+2}  \, , \nonumber 
\end{eqnarray}
\vskip -0.5 cm
\beq
c^\pm_j = -i \, \delta u \, \pm \frac{a-b}{2} - j + 1 \, , \quad d_j =\frac{a + 1 - j}{2} \, , \quad \tilde d_j = \frac{b + 1 - j}{2} \,  , \label{defX}
\eeq
where $\delta u \, = \, u - v$ and
\beq
D \, = \, \frac{x^- -y^+}{x^+ - y^-} \sqrt{\frac{x^+}{x^-}} \sqrt{\frac{y^-}{y^+}} 
\eeq
simply applying the $3\gamma, 1\gamma$ transformation
\beq
x^- \, \rightarrow \, 1/x^- \, , \qquad y^+ \, \rightarrow \, 1/y^+ \, .
\eeq
At lowest order in $g$,
\beq
D \, = \, - \frac{u^- -v^+}{u^+ - v^-} \frac{\sqrt{u^+ u^-} \sqrt{v^+ v^-}}{u^- v^+} \, .
\eeq
Including the \emph{mirror measure} \cite{BKV} for the propagation of either particle over an edge of width zero --- so when no propagators run along the common edges of the hexagons --- we obtain the expression
\beq
I(X) \, = \, \sum_{a,b = 1}^\infty \sum_{k,l = 0}^{a-1,b-1} \, \int \frac{du \, dv\, a \, b \, g^4}{4 \pi^2 (u^2 + \frac{a^2}{4})^2 (v^2 + \frac{b^2}{4})^2} \, W_1 \, W_2 \, \Sigma^{ab} \, X_k^{k,l} \, .\nonumber
\eeq
Naively, this is not a one-loop contribution, because there is a factor $g^4$ from the measure for the two bound states. Yet, we expect the scattering of the scalar constituents to introduce braiding factors like
\beq
e^{i \frac{p_1}{2}} \, e^{-i \frac{p_2}{2}} \ \stackrel{3\gamma \, 1\gamma}{\rightarrow} \ {\frac{\sqrt{u^+ u^-} \sqrt{v^+ v^-}}{g^2}}
\eeq
where we have scaled back to the field theory convention of (\ref{defXRoot}) to meet the weak-coupling expansion. Importantly, this factor does not only adjust the power to $g^2$, but it also removes the square-root branch cuts that would render inefficient the residue theorem as a means of evaluating the integrals over the rapidities $u,v$. In \cite{shotaThiago2} an \emph{averaging prescription} for such additional braiding factors is suggested. Building on the work here presented we want to study whether this prescription is the only possible one. 

Despite of the appearance, the $X$ matrix has singularities in $\delta u$ only in the lower half-plane. Poles in $X$ can therefore be avoided simply by closing the integration contour over the upper half-plane for $u$ and the lower half-plane for $v$. Doing so, the poles $u^-, \, v^+$ from the measure can contribute. Likewise, in the numerator of the phase, $\Gamma[1 + \frac{a}{2} + i \, u]$ and $\Gamma[1 + \frac{b}{2} - i \, v]$ develop singularities. Note however, that we cannot localise both rapidities by poles from the phase:
\beq
u \, = \, i \, \left(m + \frac{a}{2}\right), \ v \, = \, - i \, \left(n + \frac{b}{2} \right) \quad \Rightarrow \quad \Gamma[1 + \frac{a+b}{2} + i (u-v) ] \, = \, \Gamma[1 - m - n]
\eeq
for $m,n \, \in \, \mathbb{N}$ so that this denominator $\Gamma$-function creates a zero in these cases. Thus at least one pole, perhaps a higher one, must come from the measure. Then, e.g. with $u \, = \, i \frac{a}{2}$,
\beq
\Sigma^{ab} \, = \, \frac{\Gamma[1] \, \Gamma[1 + \frac{b}{2} - i \, v] \, \Gamma[1 + a + \frac{b}{2} + i \, v]}{\Gamma[1+a] \, \Gamma[1 + \frac{b}{2} + i \, v]  \, \Gamma[1 + \frac{b}{2} - i \, v]}
\eeq
and therefore the term in the phase that could create a pole at $v \, = \, - i (n + \frac{b}{2})$ actually drops. In conclusion, only the poles from the measure are relevant.

\section{Integrating/summing into polylogarithms}

Substituting $u = i \frac{a}{2}, \, v = - i \frac{b}{2}$ the phase reduces to $\Sigma^{ab} \, = \, \Gamma[1+a+b] / (\Gamma[1+a] \Gamma[1+b])$ and we find the cross ratio dependence
\beq
z_1^{a-k} \, \bar z_1^k \, \bar z_2^{-b+l} \, z_2^{-l}
\eeq
in accordance with the domain of convergence of the integrals over $u,v$. In order to unclutter the notation and to be able to straightforwardly Taylor-expand results in small quantities we will relabel the variables as
\beq
\left\{z_1, \, \bar z_1, \, \frac{1}{\bar z_2}, \, \frac{1}{z_2} \right\} \quad \mapsto \quad \{ z_1, \, b_1, \, y_2, \, a_2 \} \, .
\eeq
Due to the numerator of the $A$ factor in $X$, the integrand of $I(x)$ has the pole structure
$1/((u^-)^2 v^+) - 1/(u^- (v^+)^2)$. In fact, the polylog level is set by the power of these poles. In the case at hand we obtain homogeneous transcendentality 2. The residue at either double pole can create a single $\log(b_1 \, z_1)$ or $\log(a_2 \, y_2)$, respectively, as is expected from one-loop Feynman graphs, while the remaining terms are regular when all four variables become small. Derivatives from the double poles can fall onto $u^+, v^-$ creating an extra $a$ or $b$ in the denominator or onto $\delta u$ in the $m$-sum in $X$, c.f. (\ref{defX}). Due to explicit definition of $X$ we can analytically evaluate all contributions to $I(X)$ by the methods developed below on a simpler example.

In (\ref{sl2BoundBits}) we list the four types of bound states forming the complete multiplet. The $S$-matrix will then have 16 diagonal elements. There is a second instance of $X$ for the scattering of two bound states of the first type given in (\ref{sl2BoundBits}), but with $i=2$. This matrix is algebraically equal. Second, scattering of the first type of bound state over the other two is called $Y$ in \cite{glebBound} --- there are four diagonal elements for both cases, $i\, = \, 1,2$ and again, the two $Y$-matrices are equal. Last we have  six diagonal elements $Z_{jj}$ for the scattering of the last two types of bound states. Note that to some extent an averaging over braiding factors is automatic, if the dressing by momentum factors is tied to the $i$-index in $Y$; the set of $Z$-elements is symmetric in this respect. 

The $Y,Z$ cases are linear combinations of several instances of $X$ with shifted $k,l,n$-indices, with coefficients depending on $x^\pm,y^\pm$ and the counters. All unphysical poles from the coefficient matrices cancel, but this property is not manifest in the formulae spelled out in \cite{glebBound}. Ultimately, all the diagonal $Y,Z$ cases --- at least to leading order in $g$ in the given kinematics --- share the property of $X$ to have poles only in $\delta u$ in the lower half-plane. Our reasoning about the locus of poles therefore directly carries over.

In the $3\gamma, \, 1\gamma$ kinematics, the $Y$ elements are of order $1/g$. Nicely, to eliminate a single square root branch cut we also need one additional braiding factor $e^{\pm i p/2}$. Similarly, the $Z$ elements start to come in at $1/g^2$ and $Z_{11} \ldots Z_{44}$ require no braiding, while $Z_{55}, Z_{66}$ could be dressed by both, $e^{\pm i (p_1 + p_2)/2}$. Here the averaging of \cite{shotaThiago2} means to put in both possibilities with coefficient 1/2.

Remarkably, at the point $u \, = \, i \frac{a}{2}, \, v \, = \, - i \frac{b}{2}$, the matrix elements $Y_{11}, \, Y_{22}$ and  $Z_{11}, \, Z_{22}, \, Z_{33}, \, Z_{55}, \, Z_{66}$ factor into simple products of $\Gamma$-functions. Below we describe how this enables us to calculate the part of the one-loop contribution where a derivative from the residue at $1/(u^-)^2$ falls upon $(z_1 \bar z_1)^{- i u}$, or upon $1/u^+$, creating an extra factor $1/a$ (equivalently for the second particle with $z_2,v,b$). Since there is no fully explicit writing for $Y,Z$ we cannot yet compute the contribution with a derivative on the scattering matrix itself in the way described in the following, because this destroys the factorisation properties. Yet, integrating all available pieces we could pin down the space of functions and use the symbol to fit the remaining parts, also for the non-factoring cases. This could be done separately for the individual contributions. 

Let us illustrate the idea on the example of $Y_{11}$, in particular the contribution in which a derivative acts on $(z_1 b_1)^{-i u}$:
\beq
\frac{I_{\log}(Y_{11})}{\log(z_1 b_1)} \, = \, \sum_{a,b,k,l} z_1^{a-k} \, b_1^k \, y_2^{l-b} \, a_2^l  \, \frac{\quad \Gamma[a-k+b-l] \qquad  \Gamma[1+k+l]}{4 \, a \, \Gamma[a-k] \, \Gamma[1 + b - l] \ \Gamma[1+k] \,  \, \Gamma[1+l]} \label{ylSer}
\eeq
where $a,b \, = \, 1 \ldots \infty, \, k,l \, = 0 \ldots a-1,b-1$. Define
\beq
r^2 \, = \, z_1 b_1 \, , \quad p^2 \, = \, \frac{z_1}{b_1} \qquad \Rightarrow \qquad r \, \frac{\partial}{\partial r} \, z_1^{a-k} b_1^k \, = \, a \; z_1^{a-k} b_1^k \, .
\eeq
The inverse operation is $\int dr/r$. Comparing to the original series the constant part of the indefinite integral must be subtracted. We can thus eliminate $a$ from the denominator of (\ref{ylSer}) with no loss of information. Next we swap the sums over $a,k$ and $b,l$ respectively and shift the variables by $a \mapsto a + k, \, b \mapsto b + l$ in order to decouple the summations. The sums are of geometric type and yield
\beq
\frac{I_{\log}(Y_{11})}{\log(z_1 b_1)} \, = \, \int \frac{dr}{r} \frac{z_1 \, (a_2 + y_2 - a_2 y_2 - b_1 y_2 - a_2 z_1)}{
 4 \, (1 - b_1) (1 - a_2 - b_1) (1 - z_1) (1 - y_2 - z_1)} \, .
\eeq
Hence the root of the procedure is a rational function and we add polylogarithm levels by the integration in the modulus $r$:
\begin{eqnarray}
&& \frac{I_{\log}(Y_{11})}{\log(z_1 b_1)} \, = \, \frac{z_1 \, (\log[1 - b_1] - \log[1 - z_1])}{4 \, (b_1 - z_1)} + \label{oneResult} \\ && \quad \frac{z_1 \, (\log[1 - a_2] - \log[1 - a_2 - b_1] - \log[1 - y_2] + 
   \log[1 - y_2 - z_1])}{4 \, (b_1 - z_1 - b_1 y_2 + a_2 z_1)} \nonumber
\end{eqnarray}
upon subtraction of the constant part in $r$. To obtain the contribution from the derivative falling onto $1/u^+$ we can use the operation $\int dr/r$ a second time. As in \cite{shotaThiago2} we assumed that factors like $|u|, |v|$ do not arise from the $x(u)$ functions or the expansion of $\Sigma^{ab}$, and that $(-1)^{a b}$ is unphysical and has to be undone by the contraction prescription on the central hexagon.

For the other factoring matrix elements we proceed similarly. For $X$ one can again decouple the (five) sums by swapping the order of summation and shifting the counters. Since everything is expressed in terms of $\Gamma$-functions we can even address the contributions in which a single derivative falls on $\delta u$. The sums are of the type $_2F_1$ or $_3F_2$ and can be rewritten in terms of parametric integrations over Euler kernels by the standard formulae. Putting aside the integrations as long as possible we can find a path through the computation that always closes on the same type of summand/integrand. The final parametric integrations yield polylogarithms much as the integration in the modulus in the simple case above. Note that the Gauss hypergeometric function also appeared in the context of re-summing the POPE at one loop, see \cite{vanHippel} and references therein.

In our problem we find eight different denominators:
\begin{eqnarray}
&& a - y, \ b - z, \ b \ y - a \, z, \, a - y + b \, y - a \, z, \, b - b \, y - z + a \, z, \nonumber \\
&& a \, b - a \, b \, y - a \, b \, z - y \, z + a \, y \, z + b \, y \, z, \\
&& a \, b - a \, b \, y - y \, z + a \, y \, z, \ a \, b - a \, b \, z - y \, z + b \, y \, z \nonumber
\end{eqnarray}
Confusion with the bound state counters cannot arise anymore so that we dropped the $1,2$ subscript on the variables. The complete amplitudes are written as sums of weight two logarithmic functions over these denominators, with single terms of each denominator as a coefficient. Formula (\ref{oneResult}) illustrates what we mean here.

The symbol letters are
\beq
1 - a, \ a, \ 1 - b, \ 1 - a - b, \ b, \ 1 - y, \ a - y, \ y, \ 1 - z, \ b - z, \ 1 - y - z, \ z
\eeq
and
\beq	
a - y + b \, y - a \, z, \ b - b \, y - z + a \, z,  \ a \, b - a \, b \, y - a \, b \, z - y \, z + a \, y \, z + b \, y \, z 
\eeq
All the denominators are point permutations of the denominator $z-b$ of the Bloch-Wigner dilogarithm. Three of them also occur in the symbols. One can generate all the symbols of this type from
\beq
\mathrm{Li}_2\left(1 - \frac{z}{b} \right) - \frac{1}{2} \log^2(b) \label{wrongFunc}
\eeq
by permutations. 
%It is impossible to eliminate (\ref{wrongFunc}) from a sum of the amplitudes of one double glueing process. Our result from the $S$-matrix in the symmetric representation  therefore does not agree with \cite{shotaThiago2}.
At one loop, ${\cal N} = 4$ field theory results contain only Bloch-Wigner dilogarithms.
Since there will be several double glueing processes in complete correlators our results can correctly reproduce one-loop field theory if (\ref{wrongFunc}) cancels. A difficulty is that each of the three incarnations of the function occurs with various denominators making it hard to unambiguously associate terms to Bloch-Wigner dilogs or the part that has to drop.

Finally, as done in \cite{shotaThiago2}, one might choose to bring one of the particles around the central hexagon, so instead of scattering $X(u^\gamma) Y(v^{-\gamma})$ one studies $-\bar Y(v^{5 \gamma}) X(u^\gamma)$, where in this context $X,Y$ are some bound states. Due to the odd number of crossing transformations the sign of the rapidities in the phase and the $S$-matrix in the antisymmetric representation is not aligned in this version of the computation. What is more, the scalar factor $h$ contains the pole $1/(u^+ - v^+)$ now, which we would have dubbed unphysical above. Picking the residue $u \, = i \frac{a}{2}$ from the measure this becomes a pole at $v \, = \, -i (\frac{b}{2}-a)$, which is (on the border of) the lower half-plane if $b \geq 2 a$. Preliminary studies suggest that both effects introduce polylogarithms with root arguments\footnote{We thank C.~Duhr for some test calculations.}, and that uniform transcendentality may not be manifest. Yet, the end result must agree. In scattering processes with more virtual particles it will be hard to avoid this situation so that we should try to develop methods appropriate also in this kinematics. At the new residue all the factorisation properties are spoiled so that we could not proceed as before without substantial progress on simplifying $Y,Z$.

\section{Conclusions}

In the evaluation of $n$-point functions in ${\cal N}=4$ super Yang-Mills theory by hexagon tesselations, the first complicated process is the glueing of three adjacent tiles by two single mirror magnons. On the central tile this necessitates the evaluation of diagonal scattering of two so-called $sl(2)$ bound states.
% In \cite{dressCross} it is suggested that the bound state $S$-matrix in the symmetric representation in mirror/mirror kinematics yields the relevant scattering matrix, because the mirror transformation exchanges so-called $su(2)$ and $sl(2)$ sectors of the spectrum problem in ${\cal N}=4$ super Yang-Mills theory.

We find a beautiful and efficient integration scheme for this two-magnon problem, although we cannot yet ascertain that the outcome is the physical result. To answer this question must be one aim of future work.

Remarkably, the problem yields a multilinear alphabet of letters in the symbol of the relevant generalised polylogarithms, suggesting that the two-magnon problem can be integrated in closed form also beyond the leading order in the coupling constant\footnote{We thank O.~Schnetz for a discussion on this point.}.

Last, another direction of future research will be to simplify the bound state scattering matrix in the various kinematical regimes in order to be able to address higher processes, too.
%In particular, in view of \cite{dressCross} it remains to clarify why the symmetric and the antisymmetric representation (at least in $3\gamma, \, 1\gamma$ kinematics and to lowest order in the coupling) are seemingly related simply by flipping the sign of the rapidity variables.

%Nevertheless, it is argued in \cite{stringS} that the mirror rotation automatically exchanges the two sectors. As an indication, the solitonic solution known as \emph{giant magnon} \cite{hofMald} is shown to migrate from the $su(2)$ to the $sl(2)$ sector.

%\bibitem{stringS}
G.~Arutyunov and S.~Frolov,
JHEP {\bf 0712} (2007) 024 [arXiv:0710.1568 [hep-th]].
 %%CITATION = doi:10.1088/1126-6708/2007/12/024;%%

\bibitem{dressCross}
G.~Arutyunov and S.~Frolov,
J.\ Phys.\ A {\bf 42} (2009) 425401 [arXiv:0904.4575 [hep-th]].
%%CITATION = doi:10.1088/1751-8113/42/42/425401;%%

\bibitem{glebBound} 
G.~Arutyunov, M.~de Leeuw and A.~Torrielli,
Nucl.\ Phys.\ B {\bf 819} (2009) 319 [arXiv:0902.0183 [hep-th]].
%%CITATION = doi:10.1016/j.nuclphysb.2009.03.024;%%

\bibitem{psu22}
N.~Beisert,
Adv.\ Theor.\ Math.\ Phys.\  {\bf 12} (2008) 945 [hep-th/0511082].

\bibitem{cushions}
B.~Eden and A.~Sfondrini,
JHEP {\bf 1710} (2017) 098 [arXiv:1611.05436 [hep-th]].

\bibitem{shotaThiago1}
T.~Fleury and S.~Komatsu,
JHEP {\bf 1701} (2017) 130 [arXiv:1611.05577 [hep-th]].

\bibitem{shotaThiago2}
T.~Fleury and S.~Komatsu,
JHEP {\bf 1802} (2018) 177 [arXiv:1711.05327 [hep-th]].
%%CITATION = doi:10.1007/JHEP02(2018)177;%%

\bibitem{hofMald}
D.~M.~Hofman and J.~M.~Maldacena,
J.\ Phys.\ A {\bf 39} (2006) 13095 [hep-th/0604135].
%%CITATION = doi:10.1088/0305-4470/39/41/S17;%%

\bibitem{vanHippel}
H.~T.~Lam and M.~von Hippel,
JHEP {\bf 1612} (2016) 011 [arXiv:1608.08116 [hep-th]].
%%CITATION = doi:10.1007/JHEP12(2016)011;%%

\end{document}